\documentclass[11pt,a4paper]{article}
\usepackage[hyperref]{naaclhlt2018}
\usepackage{times}
\usepackage{latexsym}
\usepackage{url}
\usepackage{booktabs}
\usepackage{xfrac}
\usepackage{enumitem}
\usepackage{subcaption}
\usepackage[outline]{contour}

\usepackage{tikz}
\usetikzlibrary{arrows}
\usetikzlibrary{decorations.text}
\usetikzlibrary{calc}
\usetikzlibrary{math}
\usetikzlibrary{backgrounds}
\usetikzlibrary{shapes.misc, positioning}

\def\centerarc[#1](#2)(#3:#4:#5)
    { \draw[#1] ($(#2)+({#5*cos(#3)},{#5*sin(#3)})$) arc (#3:#4:#5); }

\aclfinalcopy 

\title{Unsupervised Keyphrase Extraction with Multipartite Graphs}

\author{Florian Boudin \\
        LS2N, Universit\'e de Nantes, France \\
        {\tt florian.boudin@univ-nantes.fr}}

\date{}

\begin{document}

\maketitle

\begin{abstract}
We propose an unsupervised keyphrase extraction model that encodes topical information within a multipartite graph structure.
Our model represents keyphrase candidates and topics in a single graph and exploits their mutually reinforcing relationship to improve candidate ranking.
We further introduce a novel mechanism to incorporate keyphrase selection preferences into the model.
%
Experiments conducted on three widely used datasets show significant improvements over state-of-the-art graph-based models.
\end{abstract}

\section{Introduction}

Recent years have witnessed a resurgence of interest in automatic keyphrase extraction, and a number of diverse approaches were explored in the literature~\cite{kim-EtAl:2010:SemEval,hasan-ng:2014:P14-1,Keyphrase:2015,augenstein-EtAl:2017:SemEval}.
Among them, graph-based approaches are appealing in that they offer strong performance while remaining completely unsupervised.
These approaches typically involve two steps: 1)~building a graph representation of the document where nodes are lexical units (usually words) and edges are semantic relations between them; 2)~ranking nodes using a graph-theoretic measure, from which the top-ranked ones are used to form keyphrases.


Since the seminal work of~\citet{mihalcea-tarau:2004:EMNLP}, researchers have devoted a substantial amount of effort to develop better ways of modelling documents as graphs.
Most if not all previous work, however, focus on either measuring the semantic relatedness between nodes~\cite{wan-xiao:2008:PAPERS,tsatsaronis-varlamis-norvrag:2010:PAPERS} or devising node ranking functions~\cite{tixier-malliaros-vazirgiannis:2016:EMNLP2016,florescu-caragea:2017:Long}.
So far, little attention has been paid to the use of different types of graphs. 
Yet, a key challenge in keyphrase extraction is to ensure topical coverage and diversity, which are not naturally handled by graph-of-words representations~\cite{hasan-ng:2014:P14-1}.


Most attempts at using topic information in graph-based approaches involve biasing the ranking function towards topic distributions~\cite{liu-EtAl:2010:EMNLP1,zhao-EtAl:2011:ACL-HLT2011,zhang-huang-peng:2013:IJCNLP}.
Unfortunately, these models suffer from several limitations: they aggregate multiple topic-biased rankings which makes their time complexity prohibitive for long documents\footnote{Recent work showed that comparable results can be achieved by computing a single topic specificity weight value for each word~\cite{Sterckx:2015:TWI:2740908.2742730,teneva-cheng:2017:Short}.}, they require a large dataset to estimate word-topic distributions that is not always available or easy to obtain, and they assume that topics are independent of one another, making it hard to ensure topic diversity.
%
For the latter case, supervised approaches were proposed to optimize the broad coverage of topics~\cite{bougouin-boudin-daille:2016:COLING,Zhang:2017:MKE:3132847.3132956}.



\begin{figure*}[!htb]
\centering

 \resizebox{\textwidth}{!}{%
  \input{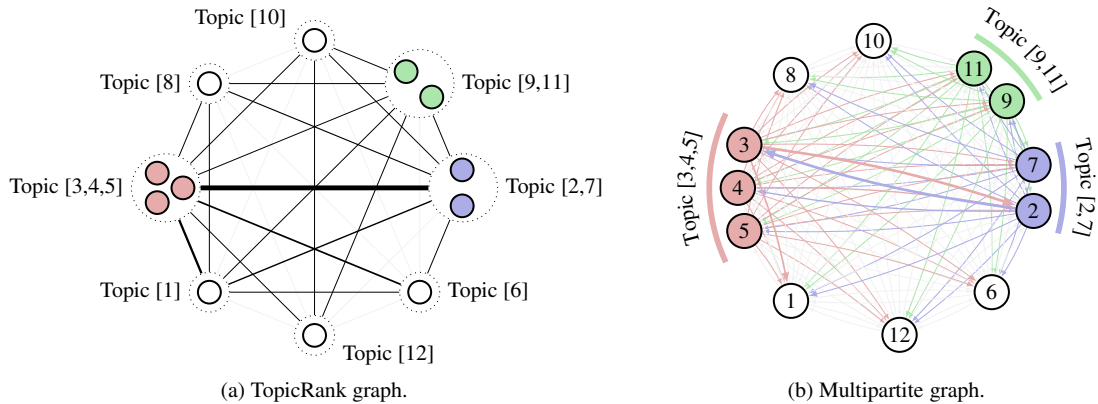}
  \begin{subfigure}{0pt}
      \phantomsubcaption{}
      \label{fig:topicrank}
  \end{subfigure}
  \begin{subfigure}{0pt}
      \phantomsubcaption{}
      \label{fig:multipartite}
  \end{subfigure}
  }
 
\caption{Comparison between TopicRank~\cite{bougouin-boudin-daille:2013:IJCNLP} and our multipartite graph representation for document 2040.abstr from the Hulth-2003 dataset. Nodes are topics (left) or keyphrase candidates (right), and edges represent co-occurrence relations.}
\label{fig:vs-graph}
\end{figure*}

Another strand of work models documents as graphs of topics and selects keyphrases from the top-ranked ones~\cite{bougouin-boudin-daille:2013:IJCNLP}.
This higher level representation (see Figure~\ref{fig:topicrank}), in which topic relations are measured as the semantic relatedness between the keyphrase candidates they instantiate, was shown to improve the overall ranking and maximize topic coverage.
%
%
The downside is that candidates belonging to a single topic are viewed as equally important, so that post-ranking heuristics are required to select the most representative keyphrase from each topic.
Also, errors in forming topics propagate throughout the model severely impacting its performance.

Here, we build upon this latter line of work and propose a model that implicitly enforces topical diversity while ranking keyphrase candidates in a single operation.
To do this, we use a particular graph structure, called multipartite graph, to represent documents as tightly connected sets of topic related candidates (see Figure~\ref{fig:multipartite}).
This representation allows for the seamless integration of any topic decomposition, and enables the ranking algorithm to make full use of the mutually reinforcing relation between topics and candidates.


Another contribution of this work is a mechanism to incorporate intra-topic keyphrase selection preferences into the model.
It allows the ranking algorithm to go beyond semantic relatedness by leveraging information from additional salience features.
%
Technically, keyphrase candidates that exhibit certain properties, e.g.~that match a thesaurus entry or occur in specific parts of the document, are promoted in ranking through edge weight adjustments.
Here, we show the effectiveness of this mechanism by introducing a bias towards keyphrase candidates occurring first in the document.

\section{Proposed Model}
\label{sec:proposed-model}

Similar to previous work, our model operates in two steps.
We first build a graph representation of the document (\S\ref{subsec:multipartite}), on which we then apply a ranking algorithm to assign a relevance score to each keyphrase (\S\ref{subsec:ranking}).
We further introduce an in-between step where edge weights are adjusted to capture position information (\S\ref{subsec:graph-weight-adjustment}).


For direct comparability with~\newcite{bougouin-boudin-daille:2013:IJCNLP}, which served as the starting point for the work reported here, we follow their setup for identifying keyphrase candidates and topics.
Keyphrase candidates are selected from the sequences of adjacent nouns with one or more preceding adjectives (\texttt{/Adj*Noun+/}).
They are then grouped into topics based on the stem forms of the words they share using hierarchical agglomerative clustering with average linkage.
Although simple, this method gives reasonably good results. 
There are many other approaches to find topics, including the use of knowledge bases or unsupervised probabilistic topic models.
%
%
Here, we made the choice not to use them as they are not without their share of issues (e.g.~limited coverage, parameter tuning), and leave this for future work.


\subsection{Multipartite graph representation}
\label{subsec:multipartite}

A complete directed multipartite graph is built, in which nodes are keyphrase candidates that are connected only if they belong to different topics.
Again, we follow~\cite{bougouin-boudin-daille:2013:IJCNLP} and weight edges according to the distance between two candidates in the document.
More formally, the weight $w_{ij}$ from node $i$ to node $j$ is computed as the sum of the inverse distances between the occurrences of candidates $c_i$ and $c_j$:
\begin{equation}
    w_{ij} = \sum_{p_i \in \mathcal{P}(c_i)}\ \sum_{p_j \in \mathcal{P}(c_j)} \frac{1}{|p_i - p_j|}
    \label{eq:edge-weighting}
\end{equation}
where $\mathcal{P}(c_i)$ is the set of the word offset positions of candidate $c_i$.
This weighting scheme achieves comparable results to window-based co-occurrence counts without any parameter tuning.


The resulting graph is a complete ${k}$-partite graph, whose nodes are partitioned into $k$ different independent sets, $k$ being the number of topics.
As exemplified in Figure~\ref{fig:vs-graph}, our graph representation differs from the one of~\cite{bougouin-boudin-daille:2013:IJCNLP} in two significant ways.
First, topics are encoded by partitioning candidates into sets of unconnected nodes instead of being subsumed in single nodes.
%
%
%
Second, edges are directed which, as we will see in \S\ref{subsec:graph-weight-adjustment}, allows to further control the incidence of individual candidates on the overall ranking.

The proposed representation makes no assumptions about how topics are obtained, and thus allows direct use of any topic decomposition.
It implicitly promotes the number of topics covered in the selected keyphrases by dampening intra-topic recommendation, and captures the mutually reinforcing relationship between topics and keyphrase candidates.
In other words, removing edges between candidates belonging to a single topic ensures that the overall recommendation of each topic is distributed throughout the entire graph.
Also, a benefit of encoding topic related candidates differentially is that the ones that best underpin each topic are directly given by the model.
%

\subsection{Graph weight adjustment mechanism}
\label{subsec:graph-weight-adjustment}

Selecting the most representative keyphrase candidates for each topic is a difficult task, and relying only on their importance in the document is not sufficient~\cite{hasan-ng:2014:P14-1}.
Among the features proposed to address this problem in the literature, the position of the candidate within the document is most reliable.
In order to capture this in our model, we adjust the incoming edge weights of the nodes corresponding to the first occurring candidate of each topic.

More formally, candidates that occur at the beginning of the document are promoted according to the other candidates belonging to the same topic.
Figure~\ref{fig:gwa} gives an example of applying graph weight adjustment for promoting a given candidate.
Note that the choice of the candidates to promote, i.e.~the selection heuristic, can be adapted to fit other needs such as prioritising candidates from a thesaurus.

\colorlet{lblue}{blue!50}
\colorlet{lred}{red!50}
\colorlet{vlblue}{blue!3}
\colorlet{vlred}{red!3}

\begin{figure}[!htb]
\centering
\resizebox{.32\textwidth}{!}{%

\begin{tikzpicture}

\tikzstyle{candidate}=[text centered, circle, draw, line width=1pt, minimum size=.5cm, inner sep=3pt]
\tikzstyle{arrow}=[line width=.5pt]

\draw  (20:2.5) node[candidate,fill=red!25] (B1) {3};
\draw  (0:2.5) node[candidate,fill=red!25] (B2) {4};
\draw  (-20:2.5) node[candidate,fill=red!25] (B3) {5};
\centerarc[red!25,line width=.1cm](0,0)(-25:25:3.1)
\centerarc[white,line width=0cm,postaction={decorate,decoration={text along path,raise=-.1cm,text align=center,text={Topic 2}}}](0,0)(20:-20:3.5)

\draw  (134:2.5) node[candidate,fill=blue!25] (A1) {1};
\draw  (154:2.5) node[candidate,fill=blue!25] (A2) {2};

\centerarc[blue!25,line width=.1cm](0,0)(129:159:3.1)
\centerarc[white,line width=0cm,postaction={decorate,decoration={text along path,text align=center,raise=-.1cm,text={Topic 1}}}](0,0)(154:134:3.5)




\draw[->,>=latex,vlred] (B1) to[bend left=5] (A1);
\draw[->,>=latex,vlblue] (A1) to[bend left=5] (B2);
\draw[->,>=latex,vlblue] (A1) to[bend left=5] (B3);

\draw[->,>=latex,vlblue] (A2) to[bend left=5] (B1);
\draw[->,>=latex,vlred] (B1) to[bend left=5] (A2);
\draw[->,>=latex,vlblue] (A2) to[bend left=5] (B2);
\draw[->,>=latex,vlred] (B2) to[bend left=5] (A2);
\draw[->,>=latex,vlblue] (A2) to[bend left=5] (B3);
\draw[->,>=latex,vlred] (B3) to[bend left=5] (A2);

\draw[->,>=latex,lblue] (A1) to[bend left=5] node (w13) [midway,above,inner sep=1pt] {$w_{13}$} (B1);
\draw[->,>=latex,lred] (B2) to[bend left=5] node (w41) [midway,below,inner sep=1pt,minimum size=.4cm] {$w_{41}$} (A1);
\draw[->,>=latex,lred] (B3) to[bend left=5] node (w51) [midway,below,inner sep=1pt,minimum size=.6cm] {$w_{51}$} (A1);

\draw[->,>=latex] (w41.east) to[bend right=45] node (sum) [midway, circle, draw, minimum size=.3cm,fill=white,inner sep=0pt] {\small +} (w13.east);

\draw (w51.east) to[bend right=45] (sum.south east);

\end{tikzpicture}
}

\caption{Illustration of the graph weight adjustment mechanism. Here, node $3$ is promoted by increasing the weight of its incoming edge according to the outgoing edge weights of nodes $4$ and $5$.}
\label{fig:gwa}
\end{figure}
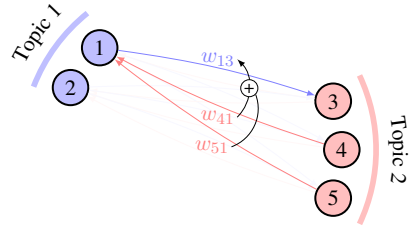

Incoming edge weights for the first occurring candidate of each topic are modified by the following equation:
\begin{equation}
w_{ij} = w_{ij} + \alpha \cdot  e^{\big(\frac{1}{p_i}\big)} \cdot \hspace{-1em} \sum_{c_k \in \mathcal{T}(c_j) \setminus{\{c_j\}}} w_{ki}
\label{math:gwa}
\end{equation}
where $w_{ij}$ is the edge weight between nodes $c_i$ and $c_j$, $\mathcal{T}(c_j)$ is the set of candidates belonging to the same topic as $c_j$, $p_i$ is the offset position of the first occurrence of candidate $c_i$, and $\alpha$ is a hyper-parameter that controls the strength of the weight adjustment.

\subsection{Ranking and extraction}
\label{subsec:ranking}

After the graph is built, keyphrase candidates are ordered by a graph-based ranking algorithm, and the top $N$ are selected as keyphrases.
Here, we adopt the widely used TextRank algorithm~\cite{mihalcea-tarau:2004:EMNLP} in the form in which it leverages edge weights:
\begin{equation}
S(c_i) = (1 - \lambda) + \lambda  \cdot \sum_{c_j \in \mathcal{I}(c_i)} \frac{w_{ij} \cdot S(c_j)}{\sum\limits_{c_k \in \mathcal{O}(c_j)}{w_{jk}}}
\label{math:textrank}
\end{equation}
where $\mathcal{I}(c_i)$ is the set of predecessors of $c_i$, $\mathcal{O}(c_j)$ is the set of successors of $c_j$, and $\lambda$ is a damping factor set to 0.85 as in~\cite{mihalcea-tarau:2004:EMNLP}.
Note that other ranking algorithms can be applied.
We use TextRank because it was shown to perform consistently well~\cite{boudin:2013:IJCNLP}.

\section{Experiments}

\begin{table*}[htb!]
    \centering
    \resizebox{\textwidth}{!}{%
    \begin{tabular}{r|rrr|rrr|rrr}
    \toprule
        ~ &
        \multicolumn{3}{c}{\textbf{SemEval-2010}} &
        \multicolumn{3}{|c}{\textbf{Hulth-2003}} &
        \multicolumn{3}{|c}{\textbf{Marujo-2012}}
    \\[.1cm]
        \textbf{Model} &
        \textbf{F$_1$@5} & \textbf{F$_1$@10} & \textbf{MAP} &
        \textbf{F$_1$@5} & \textbf{F$_1$@10} & \textbf{MAP} &
        \textbf{F$_1$@5} & \textbf{F$_1$@10} & \textbf{MAP} 
    \\
    \midrule
             
        \cite{bougouin-boudin-daille:2013:IJCNLP}
             &  9.7 & 12.3 & 7.3
             & 25.3 & 29.3 & 24.3
             & 12.1 & 17.6 & 14.6 \\
             
        \cite{Sterckx:2015:TWI:2740908.2742730}
             & 9.3 & 10.5 & 7.4
             & 21.9 & 30.2 & 25.3
             & 11.7 & 16.4 & 16.1 \\
             
        \cite{florescu-caragea:2017:Long} 
             & 10.6 & 12.2 & 8.9
             & 23.5 & 30.3 & 26.0
             & 10.9 & 17.2 & 16.1\\
    

    
    \cmidrule(l{2pt}r{2pt}){1-1}
    \cmidrule(l{2pt}r{2pt}){2-4}
    \cmidrule(l{2pt}r{2pt}){5-7}
    \cmidrule(l{2pt}r{2pt}){8-10}
    
        Proposed model
               & \textbf{12.2}$^\dagger$ & \textbf{14.5}$^\dagger$ & \textbf{11.8}$^\dagger$
               & \textbf{25.9}$^\dagger$ & \textbf{30.6} & \textbf{29.2}$^\dagger$
               & \textbf{12.5} & \textbf{18.2} & \textbf{17.2}$^\dagger$ \\
        
        w/o weight adjustment
               & 8.8 & 12.4 & 9.4
               & 21.1 & 26.8 & 25.2 
               & 12.2 & 17.8 & 16.9 \\
    \bottomrule
    \end{tabular}
    }
    \caption{F$_1$-scores computed at the top 5, 10 extracted keyphrases and Mean Average Precision (MAP) scores. $\dagger$ indicate significance at the 0.05 level using Student's t-test.}
    \label{tab:overall_results}
    \vspace*{-.2em}
\end{table*}

\subsection{Datasets and evaluation measures}

%
%

%
We carry out our experiments on three datasets:
\begin{description}[itemsep=.3em, font=\normalfont, style=standard, labelsep=0em]
    
    \item[\textbf{SemEval-2010}~\cite{kim-EtAl:2010:SemEval}], which is composed of scientific articles collected from the ACM Digital Library. We use the set of combined author- and reader-assigned keyphrases as reference keyphrases.
     
    \item \textbf{Hulth-2003}~\cite{Hulth:2003:IAK:1119355.1119383}, which is made of paper abstracts about computer science and information technology. Reference keyphrases were assigned by professional indexers.
    
    \item \textbf{Marujo-2012}~\cite{MARUJO12.672} that contains news articles distributed over 10 categories (e.g. Politics, Sports). Reference keyphrases were assigned by readers via crowdsourcing.
    

\end{description}

We follow the common practice and evaluate the performance of our model in terms of f-measure (F$_1$) at the top $N$ keyphrases, and apply stemming to reduce the number of mismatches.
We also report the Mean Average Precision (MAP) scores of the ranked lists of keyphrases.

\subsection{Baselines and parameter settings}

We compare the performance of our model against that of three baselines.
The first baseline is TopicRank~\cite{bougouin-boudin-daille:2013:IJCNLP} which is the model that is closest to ours.
The second baseline is Single Topical PageRank~\cite{Sterckx:2015:TWI:2740908.2742730}, an improved version of~\citet{liu-EtAl:2010:EMNLP1} that biases the ranking function towards topic distributions inferred by Latent Dirichlet Allocation (LDA).
%
The third baseline is PositionRank~\cite{florescu-caragea:2017:Long}, a model that, like ours, leverages additional features (word's position and its frequency) to improve ranking accuracy.

Over-generation errors\footnote{These errors occur when a model correctly outputs a keyphrase because it contains an important word, but at the same time erroneously predicts other keyphrases because they contain the same word.} are frequent in models that rank keyphrases according to the sum of the weights of their component words~\cite{hasan-ng:2014:P14-1,boudin:2015:Keyphrase}.
This is indeed the case for the second and third baselines, and we partially address this issue by normalizing candidate scores by their length, as proposed in~\cite{boudin:2013:IJCNLP}.
We use the parameters suggested by the authors for each model, and estimate LDA topic distributions on the training set of each dataset.
Our model introduces one parameter, namely $\alpha$, that controls the strength of the graph weight adjustment.
This parameter is tuned on the training set of the SemEval-2010 dataset, and set to $\alpha=1.1$ for all our experiments.
For a fair and meaningful comparison, we use the same candidate selection heuristic (\S\ref{sec:proposed-model}) across models.

\subsection{Results}

Results for the baselines and the proposed model are detailed in Table~\ref{tab:overall_results}.
Overall we observe that our model achieves the best results and significantly outperforms the baselines on most metrics.
Relative improvements are smaller on the Hulth-2003 and Marujo-2012 datasets because they are composed of short documents, yielding a much smaller search space~\cite{hasan-ng:2014:P14-1}.
%
%
%
TopicRank obtains the highest precision among the baselines, suggesting that its --one keyphrase per topic-- policy succeeds in filtering out topic-redundant candidates.
On the other hand, TopicRank is directly affected by topic clustering errors as indicated by the lowest MAP scores, which supports the argument in favour of enforcing topical diversity implicitly.
In terms of MAP, the best performing baseline is PositionRank, highlighting the positive effect of leveraging multiple features.

Additionally, we report the performance of our model without applying the weight adjustment mechanism.
Results are higher or on-par with baselines that use topic information, and show that our model makes good use of the reinforcing relations between topics and the candidates they instantiate.
We note that the drop-off in performance is more severe for F1@5 on the Semeval-2010 dataset, going from best to worst performance.
Although further investigation is needed, we hypothesise that our model struggles with selecting the most representative candidate from each topic using TextRank as a unique feature.

We also computed the topic coverage of the sets of keyphrases extracted by our model.
With over 92\% of the top-10 keyphrases assigned to different topics, our model successfully promotes diversity without the need of hard constraints. %
A manual inspection of the topic-redundant keyphrases reveals that a good portion of these are in fact clustering errors, that is, they have been wrongly assigned to the same topic (e.g.~`students' and `student attitudes').
Some exhibit a hypernym-hyponym relation while both being in the gold references (e.g.~`model' and `bayesian hierarch model' for document H-7 from the Semeval-2010 dataset), thus indicating inconsistencies in the gold data.

\section{Conclusion}

We introduced an unsupervised keyphrase extraction model that builds on a multipartite graph structure, and demonstrated its effectiveness on three public datasets.
Our code and data are available at \url{https://github.com/boudinfl/pke}.
In future work, we would like to apply ranking algorithms that leverage the specific structure of our graph representation, such as the one proposed in~\cite{becker2013ranking}.

\section*{Acknowledgements}

We thank the anonymous reviewers for their comments.
This work was supported in part by the TALIAS project (grant of CNRS PEPS INS2I 2016).
We also thank the members of the TALN team for their support.

\bibliography{biblio}
\bibliographystyle{acl_natbib}

\end{document}